\documentclass[trackchanges,twocolumn,twocolappendix,resetfootnote]{aastex701}

\usepackage{amsmath}

\begin{document}

\title{Disk-Regulated Mass Transfer Between Rotating Non-Degenerate Stars: Insights from Be and sdOB Binaries}

\author[0000-0002-0031-3029]{Zepei\,Xing}
\affiliation{Center for Interdisciplinary Exploration and Research in Astrophysics (CIERA), 1800 Sherman, Evanston, IL 60201, USA}
\email[show]{Zepei.Xing1@northwestern.edu}

\author[0000-0003-1474-1523]{Tassos\,Fragos}
\affiliation{Département d’Astronomie, Université de Genève, Chemin Pegasi 51, CH-1290 Versoix, Switzerland}
\affiliation{Gravitational Wave Science Center (GWSC), Université de Genève, CH1211 Geneva, Switzerland}
\email{Anastasios.Fragkos@unige.ch}

\author[0000-0001-9236-5469]{Vicky Kalogera}
\affiliation{Department of Physics \& Astronomy, Northwestern University, 2145 Sheridan Road, Evanston, IL 60208, USA}
\affiliation{Center for Interdisciplinary Exploration and Research in Astrophysics (CIERA), 1800 Sherman, Evanston, IL 60201, USA}
\email{vicky@northwestern.edu}

\author[0000-0001-6692-6410]{Seth\,Gossage}
\affiliation{Center for Interdisciplinary Exploration and Research in Astrophysics (CIERA), 1800 Sherman, Evanston, IL 60201, USA}
\email{seth.gossage@northwestern.edu}

\author[0000-0003-4474-6528]{Kyle~Akira~Rocha}
\affiliation{Department of Astronomy \& Astrophysics, University of California, San Diego, 9500 Gilman Drive, La Jolla, CA 92093, USA}
\email{kyrocha@ucsd.edu}

\author[0000-0002-7464-498X]{Emmanouil\,Zapartas}
\affiliation{Institute of Astrophysics, FORTH, N. Plastira 100,  Heraklion, 70013, Greece}
\affiliation{National and Kapodistrian University of Athens, 15784, Athens, Greece}
\email{ezapartas@ia.forth.gr}


\begin{abstract}

Mass transfer between non-degenerate stars is a fundamental but still poorly understood process in binary evolution. The commonly used rotationally limited accretion prescription in detailed binary evolution simulations that account for stellar rotation generally yields low accretion efficiencies that are difficult to reconcile with several observational constraints. We present a physically-motivated mass-accretion prescription in which accretion or decretion disks regulate the angular momentum transported to the accretor, thereby allowing for continued accretion at near-critical rotation. The accretion efficiency can be calculated from the conservation of the mass and the angular momentum of the disk. Analytical estimates show that the accretion efficiency depends on stellar rotation and mass ratio for direct impact accretion, and additionally on stellar radius and orbital separation in the disk accretion regime. The overall mass-weighted accretion efficiencies are close to the values expected near the threshold rotation rate, where the accreted specific angular momentum declines sharply. Applying this model to binary evolution simulations, we find that rotationally limited accretion systematically underestimates Be-star masses in Be+subdwarf O/B-type star (sdOB) systems, whereas the disk–star coupling model can produce more massive Be stars that are consistent with observations. The final binary component masses depend not only on accretion efficiency but also core-envelope mass ratio, which itself depends sensitively on the assumed overshooting. We find that our new disk-star coupling model with reduced overshooting yields component masses for Be+sdOB systems that are in closer agreement with observations.

\end{abstract}



\section{Introduction} \label{sec:intro}

Mass transfer between two non-degenerate stars is a fundamental yet insufficiently understood process within the framework of binary evolution. Accretion efficiency, which quantifies the fraction of transferred mass that can be actually retained by the accretor, is one of the key questions. It is especially critical because mass transfer frequently occurs relatively early in binary evolution to reshape the binary configuration, including the component masses and orbital period, which, in turn, determines the entire subsequent evolution. Moreover, accretors may exhibit internal structures and properties that differ from those of single stars \citep[e.g.,][]{2021ApJ...923..277R, 2023ApJ...942L..32R, 2025arXiv251220054X}.

Accretion efficiency is usually assumed to be limited by the response of the accretor to mass gain, either through radial expansion or increased rotation. Rapid mass gain could lead to significant radial inflation of the accretor forming a contact system or triggering a common-envelope phase. A commonly adopted prescription in binary population synthesis (BPS) studies is to restrict the accretion rate at $10$ times the accretor's thermal timescale accretion rate \citep{2002MNRAS.329..897H}. This criterion traces back to the analysis of \citet{1991A&A...241..419P}, based on the calculation of \citet{1977A&A....54..539K}, which showed that accreting main-sequence stars expand significantly if the thermal timescale is about $10$ times the accretion timescale. Recent studies on radial expansion of accreting stars have shown that the critical mass transfer rate that leads to significant inflation is not a single value, but depends on the stellar mass \citep{2024ApJ...966L...7L, 2024A&A...691A.174S} and the envelope structure \citep{2024MNRAS.531L..45Z}. 

In addition to responding to the added mass, stars also receive angular momentum from the transferred material. It has been suggested that even a small amount of accretion from a Keplerian accretion disk can spin up the star to its breakup limit, restricting further accretion \citep{1981A&A...102...17P}. In fact, rotationally limited accretion has been implemented in BPS studies by scaling the mass transfer rate by a factor of $1-\frac{\Omega}{\Omega_{{\mathrm{crit}}}}$ \citep{2014ApJ...796...37S}, following the approach proposed by \citet{2009MNRAS.396.1699S}, where $\Omega$ is the stellar angular velocity and $\Omega_{{\mathrm{crit}}}$ is the critical break-up value. In the widely used detailed stellar structure and evolution code \texttt{MESA} \citep{2011ApJS..192....3P,2013ApJS..208....4P,2015ApJS..220...15P,2018ApJS..234...34P,2019ApJS..243...10P, 2023ApJS..265...15J}, rotationally limited accretion can be self-consistently computed implicitly by implementing strong mass loss that ensures stars remain below critical rotation after mass transfer at each timestep \citep[see Section 2.9 of][]{2015ApJS..220...15P}. For binaries in wide orbits, rotationally limited accretion leads to a low mass-accretion efficiency. Accretion efficiency can be higher for case~A mass transfer in close orbits where tides can spin down the accretor efficiently, allowing more accretion \citep{2003astro.ph..2232L, 2012ARA&A..50..107L, 2022A&A...659A..98S, 2023ApJS..264...45F}.

Observationally, Be-star binaries provide valuable insight into mass accretion efficiency, as the rapidly rotating Be stars in binaries are believed to be the products of post-accretion spin-up \citep{2013ApJ...764..166D, 2014ApJ...796...37S,2024MNRAS.527.3076D,2025A&A...698A..38B}. \citet{2014ApJ...796...37S} found that their rotationally limited accretion leads to an overabundance of low-mass Be stars for Galactic Be X-ray binaries. A similar finding is reached by \citet{2024ApJ...971..133R}, who used the BPS code \texttt{POSYDON} \citep{2023ApJS..264...45F, 2025ApJS..281....3A}, where rotationally limited accretion is applied in the \texttt{MESA} models. Furthermore, \citet{2020MNRAS.498.4705V} suggested that the observed Be-star mass distribution in the Small Magellanic Cloud implies an accretion efficiency exceeding $\simeq 30\%$. \citet{2025A&A...704A.218X} proposed that adopting a higher accretion efficiency than that predicted by rotationally limited prescriptions may yield a better match to the large observed populations of OBe stars and Be X-ray binaries in the Small Magellanic Cloud. In recent years, a growing number of Be+subdwarf O/B-type (sdOB) star binaries have been identified \citep{2021AJ....161..248W, 2023AJ....165..203W, 2024ApJ...962...70K}. Hot subdwarfs are helium-burning stars that have been stripped of most or all of their hydrogen-rich envelopes through binary interaction and stellar winds. They are hot, compact stars that emit strongly in the ultraviolet and can be classified observationally based on their spectral features \citep{2016PASP..128h2001H, 2024arXiv241011663H}. More generally, such objects belong to the broader class of stripped-envelope helium stars, which span a continuous sequence from hot subdwarfs to Wolf–Rayet stars, depending on mass and metallicity \citep{2018A&A...615A..78G}. Because Be+sdOB binaries are the direct products of past mass-transfer phases, these systems provide particularly valuable constraints on the physics and efficiency of mass transfer. \citep{2025ApJ...987..210B} performed a population synthesis study of these systems and found that a high mass accretion efficiency ($\gtrsim0.5$) better reproduces the observations. Most recently, \citet{2025ApJ...990L..51L} used a catalog of relatively well characterized Be+sdOB binaries to estimate the required accretion efficiencies and found that half of the systems require efficiencies above $50\%$. They showed that neither rotationally limited nor thermally limited accretion prescriptions are consistent with their estimated efficiencies. 

The masses of observed binaries with Be stars indicate that extra mechanisms exist that allow more accretion in the wide, case~B mass-transfer binaries, where tidal spin down is inefficient. A possible mechanism for this is that the viscous disk coupled with the accreting star could regulate angular momentum transport \citep{1974MNRAS.168..603L, 1991ApJ...370..604P,1991ApJ...370..597P, 1991MNRAS.253...55C,1993A&A...274..796B,2013A&A...557A..40D}. \citet{1991ApJ...370..604P} found a steady-state solution for a thin accretion disk coupled with a near critically rotating accretor, where the star can continue accreting mass without gaining angular momentum and even, in some cases, accreting negative angular momentum flux. Since the solution is only for a steady state, it has been unclear how the process behaves in an evolving condition, considering both angular momentum transport and the stellar response to mass accretion. 

In this paper, we develop a simplified mass accretion model following the idea that disk-star coupling can regulate angular momentum transport, trying to gain a deeper understanding of the mass-transfer process in non-degenerate stars with both mass transfer and angular momentum transport in consideration. We use the \texttt{POSYDON} framework to run \texttt{MESA} models with our updated mass accretion prescription implemented. Then, we compare the evolution of the binaries under our new prescription with the rotationally limited model. To demonstrate the necessity of considering angular momentum transport between stars and disks, we compare the properties of our simulated Be and helium star binaries to observations. In section~\ref{sec:model}, we describe our new mass-accretion model and show analytical estimates of mass accretion efficiency. In section~\ref{sec:mesa}, we compare the evolutionary tracks of example binaries during the mass transfer between the disk-star coupling accretion model and the rotationally limited accretion model adopted in \texttt{POSYDON} binary simulation. We further discuss our findings in section~\ref{sec:diss} and present our conclusions in section~\ref{sec:conclusion}.



\section{Model assumptions} \label{sec:model}

\subsection{Angular momentum accretion}

Transferred mass flows through the inner Lagrange point $L_{1}$ may either directly hit the accretor or intersect with itself to form an accretion disk \citep{1975ApJ...198..383L}. The accretion mode depends on whether the accretor's radius $R_{\mathrm{a}}$ exceeds a minimum radius $R_{\mathrm{min}}$ \citep{1975ApJ...198..383L}, which was analytically fitted by \citet{1976ApJ...206..509U}:

\begin{equation}
R_{\mathrm{min}} \approx 0.0425 a(q+q^{2})^{0.25},
\end{equation}
where $a$ is the orbital separation of the binary, and $q$ is the mass ratio, defined as $M_{\mathrm{a}}/M_{\mathrm{d}}$, where $M_{\mathrm{a}}$ and $M_{\mathrm{d}}$ are the masses of the accretor and donor, respectively. If $R_{\mathrm{a}}<R_{\mathrm{min}}$, the transferred material forms a Keplerian disk that extends to the accretor's surface. In this case, the specific angular momentum carried by the material transported to the star is $j_{\mathrm{tr}} = \sqrt{GM_{\mathrm{a}}R_{\mathrm{a}}}$, where $G$ is the gravitational constant. If $R_{\mathrm{a}}>R_{\mathrm{min}}$, the inflow directly impacts the accretor. \citet{1975ApJ...198..383L} argued that in this regime the specific angular momentum of the accreted material is equal to the Keplerian angular momentum at the circularization radius of the stream, which is approximately $1.7R_{\mathrm{min}}$. This gives $j_{\mathrm{tr}}=\sqrt{GM_{\mathrm{a}}1.7R_{\mathrm{min}}}$. The above prescriptions have been implemented in \texttt{MESA} following \citet{2013ApJ...764..166D}. %

\begin{figure}[t]
\includegraphics{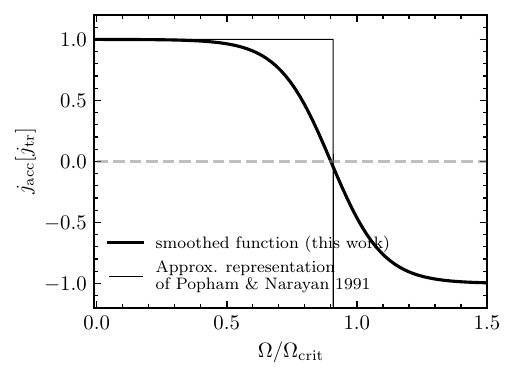}
\caption{The adopted dependence of $j_{\mathrm{acc}}$ on $\Omega/\Omega_{\mathrm{crit}}$, shown together with an approximate representation of the trend from \citet{1991ApJ...370..604P}. The dashed line denotes $j_{\mathrm{acc}}=0$.
\label{fig:jacc}}
\end{figure}

\citet{1991ApJ...370..604P} developed a steady-state thin accretion-disk model around a uniformly-rotating star using a two-dimensional fluid with a polytropic equation of state and $\alpha$-viscosity prescription. In their framework, the specific angular momentum accreted by the star was treated as a free parameter and was shown to behave like an eigenvalue that is self-consistently determined by the boundary conditions the disk satisfies. Based on the calculation of \citet{1991ApJ...370..604P}, the specific angular momentum accreted by a star, $j_{\mathrm{acc}}$, depends on the stellar rotation rate. When the ratio of the surface angular velocity to the critical value, $\Omega/\Omega_{\mathrm{crit}}$, is small, the disk's angular momentum profile must reach a maximum close to the stellar surface to connect to the slowly rotating star. At this maximum, the shear stress vanishes. Consequently, $j_{\mathrm{acc}}$ is approximately equal to the Keplerian specific angular momentum at the stellar equator. In contrast, once $\Omega/\Omega_{\mathrm{crit}}$ exceeds a certain threshold ($\sim0.9$), the disk has a continuous transition from a Keplerian rotation profile to the constant stellar rotation near the disk-star boundary. In this regime, $j_{\mathrm{acc}}$ can decrease sharply as shear stresses can move angular momentum outward from the star into the disk, allowing for continued mass accretion without the star exceeding its critical rotation limit. \citet{1991ApJ...370..604P} also emphasized that the value $\sim0.9$ should not be interpreted as a strict maximum rotation rate, owing to the ambiguous boundary between the star and the disk in their analysis, which they defined as the radius where the disk vertical half-thickness to stellar radius reaches $0.1$. Nevertheless, they argued that this choice provides a physically reasonable approximation to the stellar surface. We discuss the impact of the adopted threshold value on our calculations in Section~\ref{subsec:ama}. Although this solution is valid for a steady-state disk-star coupling, its long-term implication is that the star tends to rotate at the threshold value of $\Omega/\Omega_{\mathrm{crit}}$. 
To capture this long-term behavior while allowing a time-dependent treatment of the disk-star coupling in 1D simulations and avoiding numerical convergence issues that a discontinuous prescription of $j_{\mathrm{acc}}$ would cause, we adopt a smoother prescription in which $j_{\mathrm{acc}}$ decreases with increasing $\Omega/\Omega_{\mathrm{crit}}$. The expression we use is:
\begin{equation} \label{eq:jacc}
j_{\mathrm{acc}}/j_{\mathrm{tr}} = \frac{2}{1+e^{k(\Omega/\Omega_{\mathrm{crit}}-0.9)}}-1.
\end{equation}
We adopt $k=10$ as the default value and discuss the influence of the functional form and the value of $k$ in Section~\ref{subsec:ama}. The critical angular velocity is given by $\Omega_{\mathrm{crit}} = \sqrt{(1-L/L_{\mathrm{Edd}})~GM/R^{3}}$, where $L_{\mathrm{Edd}}=4\pi cGM/\kappa$ Here, $c$ is the speed of light and $\kappa$ is the opacity. The shape of this function and the approximate representation of the trend in $j_{\mathrm{acc}}$ from \citet{1991ApJ...370..604P} are shown in Figure~\ref{fig:jacc}. The adopted prescription scales $j_{\mathrm{acc}}$ between $-1$ and $1$ in units of $j_{\mathrm{tr}}$. It is close to $1$ at low rotation rate, decreases to $0$ at the adopted threshold rotation fraction of $0.9$, and becomes negative at higher rotation rates. The excess angular momentum, given by the difference between $j_{\mathrm{acc}}$ and $j_{\mathrm{tr}}$, is transported back into the disk. We assume that the angular momentum extraction by the disk is independent of other mechanisms of angular momentum transport, such as tides. Tidal torques are applied independently to the envelope shells. Tidal spin-down can also facilitate accretion.

For disk accretion, we adopt $j_{\mathrm{tr}}=\sqrt{GM_{\mathrm{a}}R_{\mathrm{a}}}$. In the direct impact accretion regime, once the star approaches critical rotation, we assume that a decretion disk forms that acts similarly to an accretion disk in regulating angular momentum. In this case, we adopt $j_{\mathrm{tr}}=\sqrt{GM_{\mathrm{a}}1.7R_{\mathrm{min}}}$. The same assumption of the formation of a decretion disk has been adopted by \citet{2013A&A...557A..40D}. The formation of an equatorial decretion disk induced by high rotation \citep{2011A&A...527A..84K}, together with observational and theoretical evidence for mass and angular momentum loss through a Be-star decretion disk \citep[e.g.,][]{2025MNRAS.539L..31M}, also supports this assumption.

\subsection{Angular momentum loss and accretion efficiency}

Following the prescription introduced in the previous section, in which angular momentum is returned to the disk, we assume that the excess angular momentum is transported outward through the disk and ultimately lost at its outer edge, $R_{\mathrm{out}}$. Because the Keplerian specific angular momentum at $R_{\mathrm{out}}$ is much larger than that at the stellar surface, only a small amount of mass loss at $R_{\mathrm{out}}$ is required to carry away the angular momentum extracted from the star. Then, assuming the disk remains in an equilibrium state, conservation of angular momentum gives:
\begin{equation} \label{eq:con}
j_{\mathrm{tr}} \dot{M}_{\mathrm{tr}}= j_{\mathrm{acc}} \dot{M}_{\mathrm{acc}}+j_{\mathrm{loss}} \dot{M}_{\mathrm{loss}},
\end{equation}
where $j_{\mathrm{loss}} = \sqrt{GM_{\mathrm{a}}R_{\mathrm{out}}}$. Then, we define the mass accretion efficiency $\eta \equiv\dot{M}_{\mathrm{acc}}/\dot{M}_{\mathrm{tr}}$. With the assumption of conservation of angular momentum and mass, the model permits a high accretion rate while preventing the star from exceeding the critical rotation.

In Equation~\ref{eq:con}, $j_{\mathrm{loss}}$ quantifies how efficiently the angular momentum transported into the disk can be removed from binary systems, and therefore directly determines the accretion efficiency. A larger value of $j_{\mathrm{loss}}$, or equivalently a larger disk radius $R_{\mathrm{out}}$ in our model, implies that less mass loss from the system is required to extract the residual angular momentum, resulting in a higher accretion efficiency. By default, we adopt $R_{\mathrm{out}}$ as $0.8$ Roche-lobe radius of the accretor $R_{\mathrm{RL,a}}$, corresponding to the approximate tidal truncation radius of the disk \citep{1977ApJ...216..822P}. The Roche-lobe radius is computed as \citep{1983ApJ...268..368E}:
\begin{equation} 
R_{\mathrm{RL,a}} = \frac{0.49q^{2/3}}{0.6q^{2/3}+\ln(1+q^{1/3})}a.
\end{equation}
If the accretor's radius exceeds $0.8R_{\mathrm{RL,a}}$, we further assume that the disk extends out to $R_{\mathrm{RL,a}}$.

\subsection{Analytical results}
In this section, we calculate the accretion efficiency analytically based on the proposed model. Combining Equation~(\ref{eq:jacc}) and Equation~(\ref{eq:con}), and substituting the expressions for $j_{\mathrm{tr}}$ and $j_{\mathrm{loss}}$, the accretion efficiency $\eta$ can be written, using mass conservation $\dot{M}_{\mathrm{tr}} = \dot{M}_{\mathrm{acc}}+\dot{M}_{\mathrm{loss}}$, as:
\begin{equation} 
\eta = 1- \frac{2-2/(1+e^{10(\Omega/\Omega_{\mathrm{crit}}-0.9)})}{\sqrt{R_{\mathrm{out}}/R_{\mathrm{in}}}-2/(1+e^{10(\Omega/\Omega_{\mathrm{crit}}-0.9)})+1},
\end{equation}
where $R_{\mathrm{in}} = R_{\mathrm{a}}$ for disk accretion and $R_{\mathrm{in}} = 1.7R_{\mathrm{min}}$ for direct-impact accretion. At the threshold rotation rate $\Omega/\Omega_{\mathrm{crit}}=0.9$, $\eta=1-\sqrt{\frac{R_{\mathrm{in}}}{R_{\mathrm{out}}}}$. 

\begin{figure*}[ht!]
\includegraphics{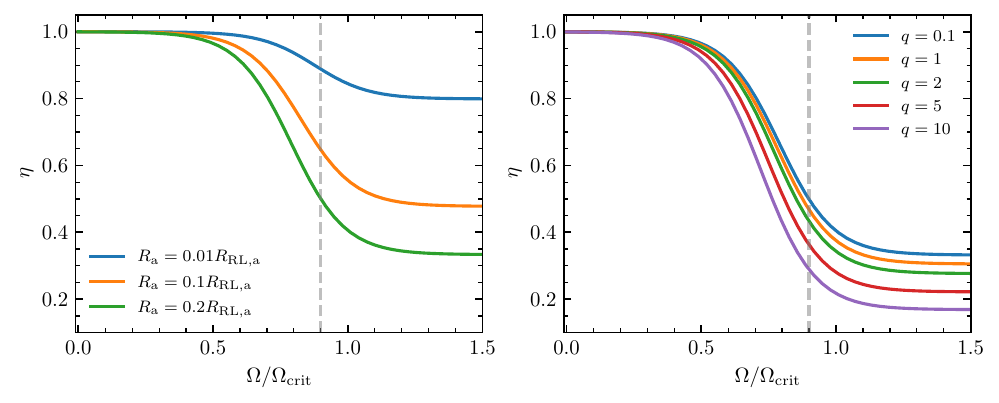}
\caption{Accretion efficiency as a function of $\Omega/\Omega_{\mathrm{crit}}$ for different accretor radii in units of the Roche-lobe radius in the disk-accretion regime (left) and for different mass ratios $q= M_{\mathrm{a}}/M_{\mathrm{d}}$ in the direct-impact regime (right). The dashed line indicates the threshold $\Omega/\Omega_{\mathrm{crit}} = 0.9$. 
\label{fig:eta}}
\end{figure*}

In the case of disk accretion, $R_{\mathrm{a}}<R_{\mathrm{min}}$ must be satisfied. Over the mass ratios interval $q\in[0.1,10]$, $R_{\mathrm{min}}/R_{\mathrm{RL,a}}$ increases from approximately $0.118$ to $0.238$. Based on this range, we evaluated $\eta$ as a function of $\Omega/\Omega_{\mathrm{crit}}$ in the disk accretion regime using three different accretor radii in units of its Roche-lobe radius, $R_{\mathrm{a}}=0.01,0.1,0.2~R_{\mathrm{RL,a}}$, which are shown in the left panel of Figure~\ref{fig:eta}. The accretion efficiency $\eta$ decreases with increasing $\Omega/\Omega_{\mathrm{crit}}$, since higher rotation rates require a larger fraction of the angular momentum to be removed from the disk, thereby increasing mass loss. In general, $\eta$ decreases with increasing accretor's radius because the specific angular momentum at the equator becomes larger. For a fixed accretor radius, $\eta$ increases with increasing mass ratio and binary separation as its Roche-lobe radius increases with both quantities. This is a consequence of the corresponding increase in disk size with larger Roche-lobe radius, which allows for more efficient angular-momentum removal. In the case of direct-impact accretion, $R_{\mathrm{out}}/R_{\mathrm{in}}$ depends only on $q$. Accordingly, $\eta$ becomes a function of $\Omega/\Omega_{\mathrm{crit}}$ and $q$. In the right panel of Figure~\ref{fig:eta}, we display the accretion efficiencies as a function of $\Omega/\Omega_{\mathrm{crit}}$ for multiple values of $q$. For all rotation rates, $\eta$ decreases with increasing $q$. At the threshold rotation value, the maximum accretion efficiency reaches approximately $0.5$ for small mass ratios. 

\section{\texttt{MESA} simulations} \label{sec:mesa}

The stellar and binary physics adopted in our MESA simulations follow the prescriptions implemented in \texttt{POSYDON} version 2.0.0, as described in \citet{2023ApJS..264...45F} and updated in \citet{2025ApJS..281....3A}, using \texttt{MESA} revision 11701 \footnote{In the subroutine that sets the rotational velocity of newly accreted material, \texttt{MESA} does not permit negative angular momentum in the two boundary layers after re-meshing. We comment out this restriction to avoid occasional interruptions when the star accretes material carrying negative angular momentum flux. We verify that, after solving the angular-momentum transport, the accretors never develop retrograde rotation.}. Here we highlight the modifications introduced in this work. 

To implement our disk-star coupling accretion model, we disable the implicit-wind scheme \citep{2015ApJS..220...15P}, used for preventing supercritical rotation, retaining only the explicit prescription for rotationally enhanced winds \citep{1998A&A...329..551L,2000ApJ...528..368H}. In this case, accretors may temporarily exceed the critical rotation due to rapid radial expansion. However, in our model, the explicit rotationally enhanced winds together with the reduced accretion rate in the supercritical regime restore the star to subcritical rotation on short timescales. Moreover, \citet{1991ApJ...370..597P} proposed that stars can accrete even while rotating above the critical rate due to nonzero pressure at the equatorial plane. 

Because we assume that the mass is lost from the outer edge of the disk rather than the vicinity of the accretor as isotropic re-emission, the orbital angular momentum loss is computed by integrating the angular momentum loss at the disk edge relative to the binary’s center of mass by default. The specific angular momentum of a mass element lost at the disk edge, averaged over the azimuthal angle $\phi$, is given by
\begin{equation} 
\begin{split}
\langle h_{\mathrm{ml}} \rangle_{\phi} = 
\langle \boldsymbol{R}_{\mathrm{COM}}\times (\boldsymbol{v}_{\mathrm{orb}}+\boldsymbol{v}_{\mathrm{disk}}) \rangle_{\phi} \\
=\frac{aM_{\mathrm{d}}}{M_{\mathrm{d}}+M_{\mathrm{a}}}|v_{\mathrm{orb}}| + R_{\mathrm{out}}|v_{\mathrm{disk}}|,
\end{split}
\end{equation}
where $\boldsymbol{R}_{\mathrm{COM}}$ is the vector from the mass element to the center of mass, $\boldsymbol{v}_{\mathrm{orb}}$ is the accretor's orbital velocity, and $\boldsymbol{v}_{\mathrm{disk}}$ is the Keplerian velocity of the disk material at the disk edge. Then, the total orbital angular momentum due to mass loss is 
\begin{equation} 
\dot{J}_{\mathrm{ml}} = \dot{M}_{\mathrm{loss}}\left( \frac{aM_{\mathrm{d}}}{M_{\mathrm{d}}+M_{\mathrm{a}}}\right)^{2}\frac{2\pi}{P_{\mathrm{orb}}} + \dot{M}_{\mathrm{loss}} \sqrt{GM_{\mathrm{a}}R_{\mathrm{out}}} .
\end{equation}

\subsection{Evolution of an example binary system} \label{subsec:individual}

\begin{figure*}[t]
\includegraphics[width=\textwidth]{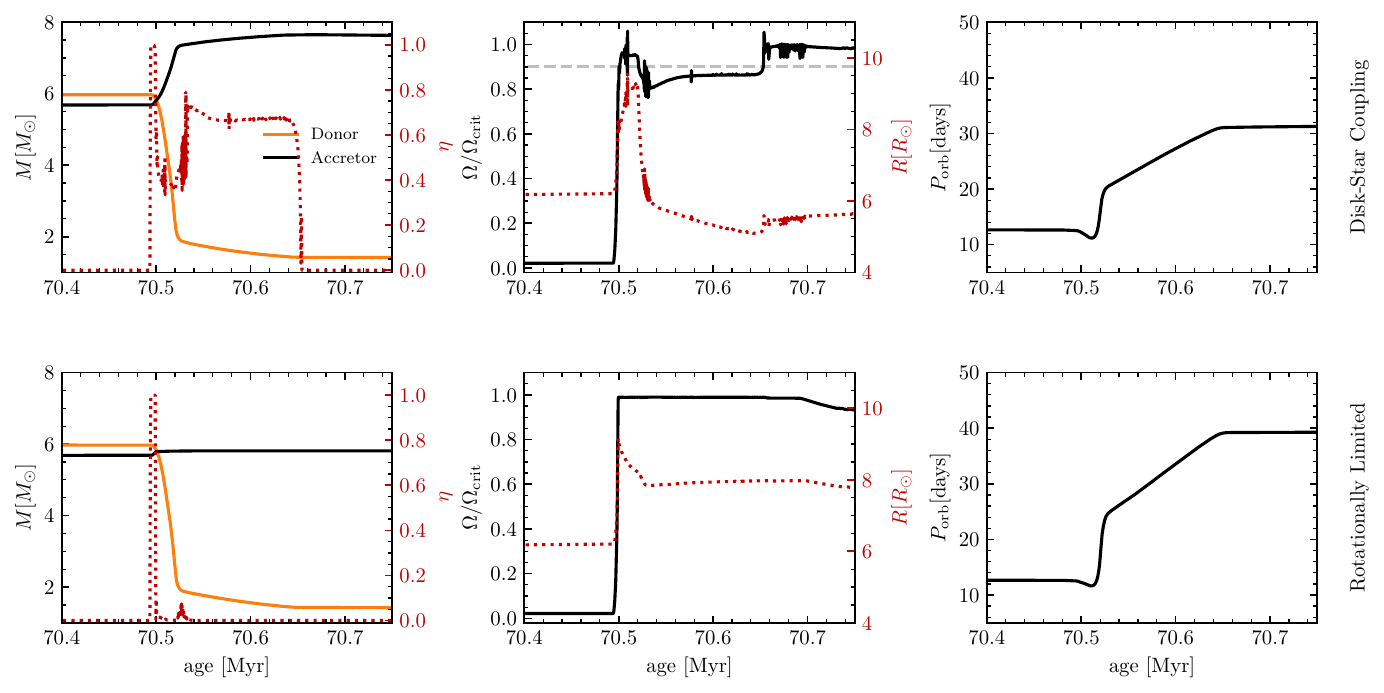}
\caption{Evolution of an example binary under the two accretion models, showing the component masses, accretion efficiency, angular velocity relative to the critical value, accretor radius, and orbital period during mass transfer. The gray dashed horizontal line in the top middle panel marks the threshold of $\Omega/\Omega_{\mathrm{crit}} = 0.9$. 
\label{fig:compare}}
\end{figure*}

First, we compare the binary evolution during mass transfer under the rotationally limited accretion prescription and under our disk–star coupling model. Figure~\ref{fig:compare} shows the evolution of an example binary for the two models, including the component masses, accretor radius, rotational velocity relative to the critical value, and orbital period. The binary initially consists of two stars with initial masses of $6\,M_{\odot}$ and $5.7\,M_{\odot}$ in a $12.6\,\mathrm{d}$ orbit, and subsequently undergoes case~B mass transfer. 

In the rotationally limited model, as the donor star transfers $\simeq 4\,M_{\odot}$ to the accretor, the accretor accretes only $\simeq 0.2\,M_{\odot}$ before reaching critical rotation, and subsequently barely accretes any additional material. Over the entire mass-transfer phase, the orbital period increases to $\simeq40\,\mathrm{d}$. In contrast, in the disk-star coupling model, the accretor gains $\simeq1.6\,M_{\odot}$ during a rapid mass-transfer phase, when the mass transfer rate exceeds $\sim 10^{-5}\,M_{\odot}\,\mathrm{yr^{-1}}$, with an accretion efficiency of $\simeq0.4$. The star accretes $\simeq0.3\,M_{\odot}$ during a subsequent slow mass transfer phase, when the mass transfer rate is below $\sim10^{-5}\,M_{\odot}\,\mathrm{yr^{-1}}$, with an accretion efficiency of $\approx0.7$. At the onset of mass transfer, the accretion timescale ($\sim10^{5}\,\mathrm{yr}$) is shorter than the thermal timescale ($\sim10^{6}\,\mathrm{yr}$) of the accretor, causing the star to expand rapidly and spin up to a high rotation rate, even exceeding the threshold. Then, it accretes negative angular momentum from the inflowing material and continues to expand. After this rapid mass transfer phase, the accretion rate decreases and the star can radiate away the excess thermal energy stored in the envelope and shrink back to a smaller radius. Due to substantial mass accretion and internal mixing, hydrogen-rich material is transported into the stellar core, increasing the core mass and leading the star to settle into a more compact configuration once thermal equilibrium is re-established. The radial contraction during thermal relaxation leads to a decrease in $\Omega/\Omega_{\mathrm{crit}}$, as noted by \citet{2026arXiv260108508W}. Consequently, the accretion efficiency increases to $\simeq0.8$. The accretor then spins up slightly due to the slow mass accretion, but $\Omega/\Omega_{\mathrm{crit}}$ remains just below the threshold, keeping the accretion efficiency at $\simeq0.7$. Although the mass ratio becomes more extreme than that in the rotationally limited model, the orbital period increases only to $\simeq30\,\mathrm{d}$ because the orbital angular momentum is more efficiently removed at the disk edge.


Since $\Omega_{\mathrm{crit}}$ is a steep function of stellar radius, the radial response of the accretor differs between the two models. In the rotationally limited accretion model, once the star first reaches critical rotation, its radius simultaneously attains its maximum. Because the transferred mass carries the Keplerian specific angular momentum at the stellar surface, the accretor can only spin up. Consequently, to maintain subcritical rotation, fast winds must remove substantial amounts of mass and angular momentum to prevent further radial expansion. In the case of disk-star coupling model, the accretor is able to accept more mass and expand gradually during accretion as long as the disk efficiently extracts the angular momentum transferred to the accretor. 

\subsection{Be and helium star binaries} \label{subsec:Be}

In this section, we compare the mass and orbital period range of our simulated Be and helium star binaries with those of the observed systems. For the observed sample, we follow \citet{2025ApJ...990L..51L} and additionally include one recently well characterized system: V742~Cas \citep{2025ApJ...995..180G}. The observed systems selected by \citet{2025ApJ...990L..51L} represent relatively well constrained Be+sdOB binaries, primarily drawn from \citet{2023AJ....165..203W} and \citet{2024ApJ...962...70K}. The sample also includes FY~CMa \citep{2008ApJ...686.1280P}, 59~Cyg \citep{2013ApJ...765....2P}, $\phi$~Per \citep{2015A&A...577A..51M}, Binary HD~55606\citep{2018ApJ...865...76C}, and $\kappa$~Dra \citep{2022ApJ...940...86K}. 

We simulate a grid of binary models at solar metallicity with primary masses in the range of $2-9\,M_{\odot}$, stepped by $0.5\,M_{\odot}$, mass ratios from $0.4$ to $0.95$ with a spacing of of $0.05$, and eleven orbital periods uniformly spaced in logarithmic space between $10^{0.2}$ and $10^{2}$ days. In addition to the termination conditions implemented in \texttt{POSYDON} \texttt{MESA} setup, we stop the simulation when the secondary reaches the terminal age main sequence. 

\begin{figure*}[t]
\includegraphics[width=\textwidth]{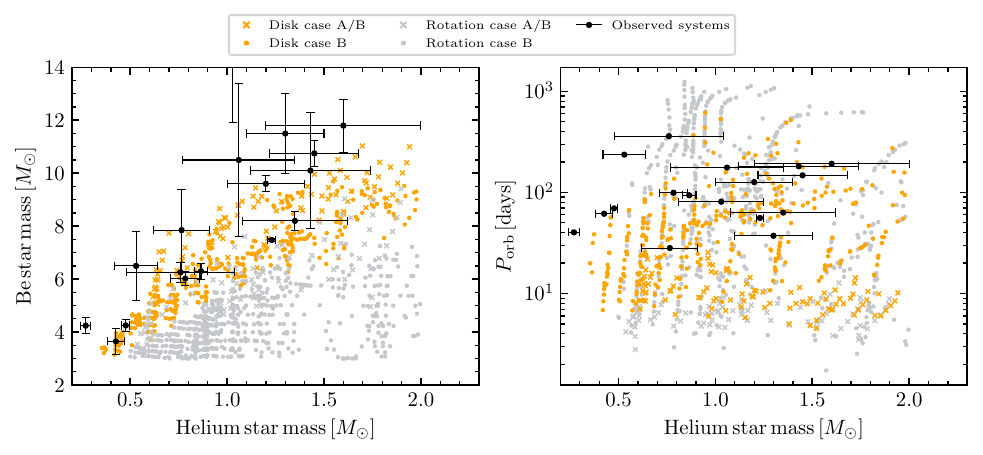}
\caption{Parameter space of component masses and orbital periods for the observed Be+sdOB systems and the simulated Be and helium star binaries under both the rotationally limited (gray) and disk–star coupling (orange) accretion models. Systems that undergo case~A mass transfer are indicated with crosses and systems undergo case~B mass transfer are shown with dots. The observed systems are primarily drawn from \citet{2023AJ....165..203W} and \citet{2024ApJ...962...70K, 2025A&A...694A.208K}, and include FY~CMa \citep{2008ApJ...686.1280P}, 59~Cyg \citep{2013ApJ...765....2P}, $\phi$~Per \citep{2015A&A...577A..51M}, HD~55606 \citep{2018ApJ...865...76C}, $\kappa$~Dra \citep{2022ApJ...940...86K}, and V742~Cas \citep{2025ApJ...995..180G}.
\label{fig:grid_default}}
\end{figure*}

Figure~\ref{fig:grid_default} presents the parameter space of the masses and orbital periods of the observed systems and our simulated Be and helium star binaries for both the rotationally limited accretion model and the disk-star coupling model. We select the simulated binaries for comparison with following criteria. First, the secondary is a Be star. The criteria we used for Be stars are that the stars are main-sequence stars, with masses $\ge3\,M_{\odot}$, consistent with the spectral classification for late-type B stars \citep{Harmanec1988BAICz}, and $\Omega/\Omega_{\mathrm{crit}}>0.3$, motivated by the wide range of subcritical rotation rates inferred for Be stars \citep[e.g.,][]{Zorec+2016A&A,Balona&Ozuyar2021ApJ}. Second, the primary stars have left hydrogen burning phase and have been stripped with a surface hydrogen abundance below $0.4$ \citep[e.g.,][]{2023ApJ...959..125G}. Low-mass helium stars ($\lesssim0.5\,M_{\odot}$) often retain higher surface hydrogen abundance because of their weak winds, whereas most of the helium stars in our models are fully stripped with surface hydrogen abundance below $0.01$. Some binaries with mass ratio close to unity do not fully strip the primary before the secondary leaves main sequence. Third, both stars are within their Roche lobe. Our goal is to investigate how different mass transfer models affect the possible mass range of the resulting Be and helium stars, as well as the orbital period distribution. The data points shown are the final timestep that satisfies the criteria for our binary models. We only show one data point for each binary because the binary properties do not change significantly after mass transfer. We apply no population or observational weighting, so the density of the data points does not represent observational likelihood. 

In Figure~\ref{fig:grid_default}, we further distinguish binaries that undergo case~A mass transfer from those that experience only case~B mass transfer. The gray points represent the rotationally limited model, where the Be-star masses are systematically below those of the observed systems. Only some binaries that go through case~A mass transfer marginally reach the lower bound of the observed Be-star masses, but these systems have short orbital periods $\lesssim10\,\mathrm{d}$, which are systematically shorter than those of the observed systems. The disk-star coupling model produces Be-stars masses covering part of the observed range, but only reaches the lower end of the massive Be stars for case~A mass-transfer binaries. The resulting orbital periods are comparable to those from the rotationally limited model. This is because, although mass loss from the disk enhances the loss of orbital angular momentum, the more conservative mass transfer reduces the total amount of mass lost and leads to more extreme reverse mass ratios,  thereby accelerating orbital expansion during the mass-transfer phase. Overall, the observed systems have long orbital periods. The most massive Be stars produced in our simulations originate from case-A mass transfer, yet their orbital periods remain shorter than those observed. We note that in the population synthesis studies of \citet{2021ApJ...908...67S} and \citet{2025ApJ...987..210B}, performed with the rapid population synthesis code BSE \citep{2002MNRAS.329..897H}, the component masses of Be+sdOB binaries can be covered by case~A mass-transfer systems even when accretion is limited by rotation. The production of massive Be stars in their models may therefore result from an overestimate of the amount of mass transferred during case~A mass transfer in rapid population synthesis codes \citep{2024MNRAS.530.3706D}. 

\begin{figure*}[t]
\includegraphics[width=\textwidth]{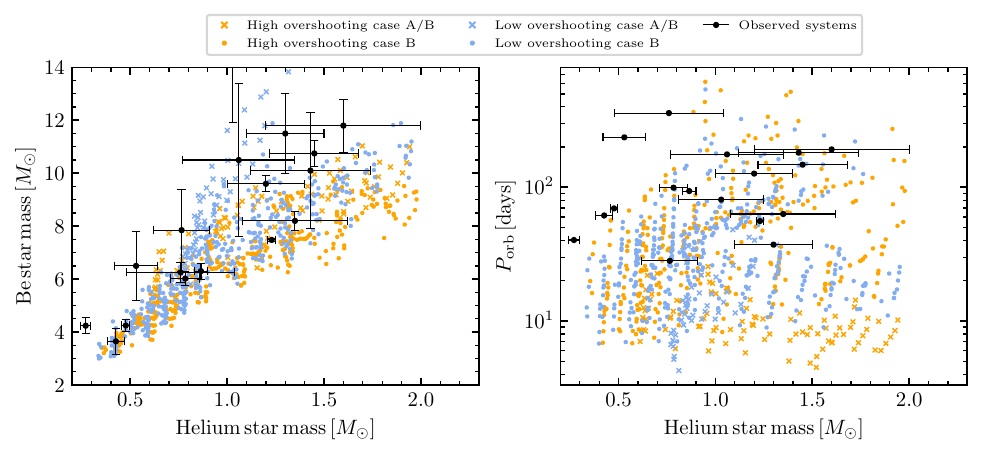}
\caption{Same as Figure~\ref{fig:grid_default} but for disk-star coupling models with two different overshooting setups.
\label{fig:grid_oveershoot}}
\end{figure*}

Overshooting plays a crucial role in this context because it directly determines the helium core mass of the primary and the envelope mass available for transfer onto the Be star. For this reason, we consider a variation in the overshooting treatment in our \texttt{MESA} simulations. In \texttt{POSYDON}, we employ the exponential decay formalism \citep{2000A&A...360..952H,2011ApJS..192....3P} with $f_{\rm ov} = 0.016$ for stars below $4\, M_\sun$ \citep{2016ApJ...823..102C} and a smooth transition to $f_{\rm ov} = 0.0415$ for stars more massive than $8\,M_\sun$. The value adopted for massive stars is converted from \citet{2011A&A...530A.115B} of $\alpha_{\rm ov} = 0.335$ with the rough ratio $\sim 10$ between the step overshoot formalism parameter $\alpha_{\rm ov}$ and the exponential decay formalism parameter $f_{\rm ov}$ \citep{2017ApJ...849...18C}. In this model variation, we adopt a reduced overshooting prescription following \citet{2021A&A...653A.144H} using the step overshoot formalism, where $\alpha_{\rm ov}$ is interpolated linearly from $0.1$ at $1.66\,M_{\odot}$ \citep{2016A&A...592A..15C} to $0.3$ at $20\,M_{\odot}$ \citep{2011A&A...530A.115B}. We plot the Be-star and the helium-star masses and the orbital periods from the disk-star coupling mass-transfer models under both overshooting prescriptions in Figure~\ref{fig:grid_oveershoot}. With lower overshooting, the helium-core masses are lower, leading to lower helium-star masses and allowing more envelope material to be transferred to the secondary, resulting in higher Be-star masses. Although the accretion efficiency is largely unaffected, the resulting component masses and orbital periods span a parameter space that more consistently matches the observations. Systems with helium-stars masses of $\sim1.5-2.0\,M_{\odot}$ and the Be star companions more massive than $\sim10\,M_{\odot}$ originate from case~A mass-transfer binaries in the high overshooting models, whereas in the low overshooting variation, this ragion of parameter space is populated by case~B mass-transfer binaries. Case~A systems still produce the most massive Be stars, but their stripped helium stars are significantly less massive. The effect of overshooting is less pronounced for low mass systems. The properties of HR6819 \citep{2025A&A...694A.208K}, which has component masses of $0.27\,M_{\odot}$ and $4.24\,M_{\odot}$, cannot be reproduced under either overshooting prescription; as argued by \citet{2026A&A...705A.225P}, even fully conservative mass transfer combined with extremely low overshooting fails to simultaneously explain its masses and orbital period.

\begin{figure}[t]
\includegraphics{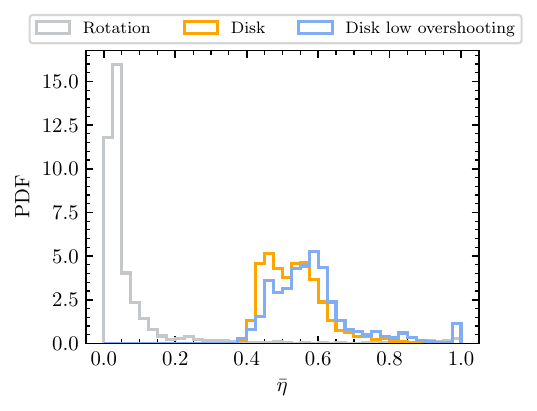}
\caption{Distribution of the mass-weighted accretion efficiency for the different models in our binary grids.
\label{fig:average}}
\end{figure}

In Figure~\ref{fig:average}, we show the distribution of the mass-weighted mass accretion efficiency following \citet{2024ApJ...971..133R} for all binaries in our grids under the different models. The mass-weighted accretion efficiency, $\bar\eta$, is defined as the sum of the total accreted mass divided by the total transferred mass over all mass transfer phases. The mass-weighted accretion efficiency is primarily determined by the phase of the most rapid mass transfer. In the rotationally limited accretion model, most binaries have accretion efficiencies below $10\%$, with a small fraction of close case~A mass-transfer systems reaching values near $100\%$. In contrast, the disk-star coupling model leads to an mass-weighted accretion efficiency above $40\%$, extending to around $60\%$. This is broadly consistent with the analytical expectation that the efficiency near the rotational threshold is around $50\%$. Reduced overshooting enhances the efficiency slightly and produces a peak around $60\%$.

\section{Discussion} \label{sec:diss}

\subsection{Angular momentum accretion} \label{subsec:ama}
Our fitted relation between stellar rotation and the specific angular momentum accreted is intended as a smoothed approximation that captures the trend of disk-regulated angular momentum extraction of \citet{1991ApJ...370..604P}, rather than a robust physical calculation. However, it is worth noting that the shape of the function, governed by the value of $k$ in Equation~\ref{eq:jacc}, has only a modest impact on the resulting accretion efficiency and angular momentum accretion, as long as the essential behavior of the accreted angular momentum as a function of the rotation rate is preserved: the accreted angular momentum is Keplerian at low rotation rates and declines to zero as the rotation rate approaches the adopted threshold. This trend of angular momentum accretion is also qualitatively consistent with the calculation of \citet{1991ApJ...370..597P}, who proposed that the specific angular momentum accreted becomes zero or even negative at critical rotation, and with \citet{1991MNRAS.253...55C}, who showed that near breakup, $j_{\mathrm{acc}}$ is proportional to $1-\frac{\Omega}{\Omega_{\mathrm{crit}}}$. The weak dependence on $k$ is because, once accretion begins, the accretor is rapidly spun up to the threshold rotation rate with little accretion, and then stays near this point due to efficient spin-down occurs whenever the threshold is exceeded. With rapid accretion, the stellar rotation rates may overshoot beyond the threshold, with the degree of overshoot depending in part on the numerical timestep. As a consequence, the angular momentum accreted, and hence accretion efficiency, is determined predominantly by the narrow regime around the threshold. Shifting the threshold also does not significantly affect the resulting accretion efficiencies, except when the threshold is set close to unity, in which case the stellar rotation rate can overshoot to values near or even exceeding unity. In this regime, the overall accretion efficiency decreases because rotationally enhanced winds become much stronger at extreme rotation rates. Ideally, the accretor's rotation rate would remain very close to the threshold if the rotational self-regulation is efficient. In Appendix~\ref{appendixb}, we show analytical estimates of the accretion efficiency and the mass-weighted accretion efficiency of the binary grids with two variations of the functional form of $j_{\mathrm{acc}}$.

The actual angular momentum exchange between the star and the disk depends on the disk model. The calculations of \citet{1991ApJ...370..604P} assumed a steady-state thin disk, whereas different disk structures and physical conditions can lead to different angular momentum transport efficiencies. Additional uncertainties such as the presence of magnetic fields and turbulence could also influence the coupling between the star and the disk. If angular momentum extraction by the disk is not sufficiently efficient, the star would spin up to critical rotation and accretion would be suppressed. These uncertainties could suggest that the accretion efficiency is inherently model dependent and also imply that binary configurations with differing disk properties may naturally exhibit different efficiencies.

\subsection{Angular momentum loss}
In our model, we assume that the excess angular momentum extracted from the star is transported through the disk and removed via mass loss at the disk’s outer edge. The mass and angular-momentum conservation within the disk links the angular momentum exchange at the inner boundary with the stellar surface to the mass loss occurring at the outer edge. This assumption requires that the angular momentum transport within the disk is efficient and the viscous timescale is shorter than the relevant stellar evolution timescale. This assumption may break down during extremely rapid mass transfer phase or strong disk perturbations, but should be adequate for describing the long-term secular evolution.

In Equation~\ref{eq:con}, any additional disk angular-momentum loss mechanisms would appear as an extra term, in addition to the contribution from mass loss. For instance, if tidal torques efficiently extract angular momentum from the disk near the tidal truncation area and transfer it to the orbit, less mass loss is required at its outer edge to satisfy angular-momentum conservation. Therefore, this would allow a higher mass accretion efficiency. However, the efficiency of this process, as well as its dependence on disk properties and the binary configuration, remains highly uncertain. As an illustrative case, we consider an extreme assumption that tides remove half of the disk’s angular momentum, then, the accretion efficiency at the rotational threshold would increases from $1- \sqrt{\frac{R_{\mathrm{out}}}{R_{\mathrm{in}}}}$ to $1- 0.5\sqrt{\frac{R_{\mathrm{out}}}{R_{\mathrm{in}}}}$. At the same time, the reduced orbital angular momentum loss together with orbital angular momentum transferred back to the orbit by tides, are expected to lead to wider binary orbits.
\begin{figure*}[t]
\includegraphics[width=\textwidth]{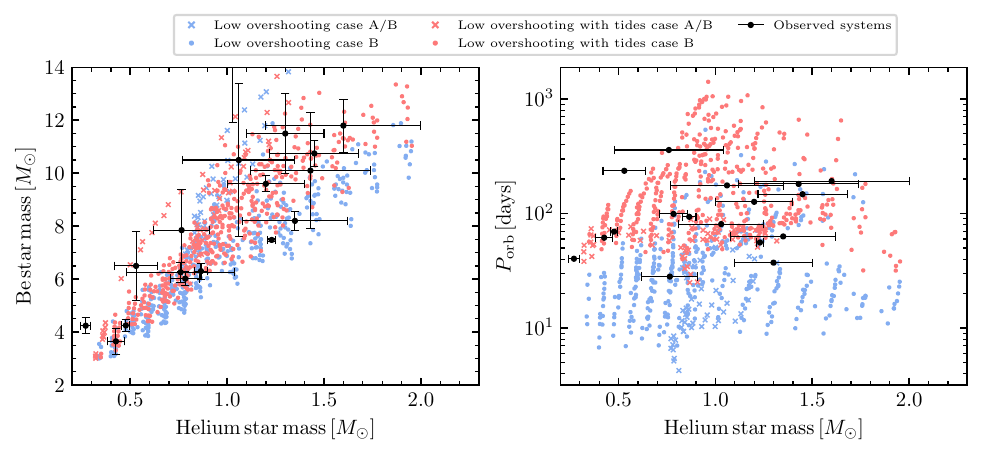}
\caption{Same as Figure~\ref{fig:grid_default} but for disk–star coupling models with low overshooting, shown both with and without tidal extraction of disk angular momentum into the orbit.
\label{fig:grid_tides}}
\end{figure*}
In Figure~\ref{fig:grid_tides}, we show the Be-star and helium-star masses and orbital periods for disk–star coupling models under low overshooting setup, both with and without including tidal extraction of disk angular momentum into the orbit. With reduced mass loss, the models that include tidal extraction can produce more massive Be stars. The impact on orbital periods is substantial. Assuming that tides extract half of the disk angular momentum, most binaries end up with orbital periods longer than $\simeq50\, \mathrm{d}$. This demonstrates that the fraction of disk angular momentum being removed from the system has a significant effect on the final orbital periods.

In our model, we assume that the mass is lost from the disk edge as fast winds and do not explicitly account for the possibility that it may instead escape through $L_{2}$ eventually. If so, the escaping material would remove orbital angular momentum more efficiently. Nevertheless, this does not alter our underlying assumption of angular momentum conservation within the disk as long as the mass is lost from the outer edge of the disk and removes the Keplerian specific angular momentum. 

\subsection{Radial expansion}

With an increased accretion efficiency compared to the rotationally limited accretion model, the instantaneous accretion timescale in the disk-star coupling model can become much shorter than the local thermal timescale of the accretor, leading to rapid radial expansion. As a result, the model produces more contact binaries than the rotationally limited model, which restricts radial expansion once the accretor reaches critical rotation. In \texttt{POSYDON}, we consider unstable mass transfer is initiated when a binary becomes contact and either binary components have left the main sequence. Under this criterion, more binaries would undergo unstable mass transfer due to the stronger radial response of the accretors. In our grid under the rotationally limited accretion model, only $\sim 6\%$ of binaries end up as unstable due to this contact condition, whereas this fraction increases to $\sim39\%$ in the disk star coupling model. However, the mass-transfer stability boundary from our model differs from that obtained in models that consider only the accretor’s thermal limit. Our model includes rotation, in which high spin rates can suppress accretion and thereby allow the accretor to adjust thermally.

\subsection{Uncertainties in direct impact accretion and disk accretion}

The amount of angular momentum transferred to the accretor in the direct-impact regime is uncertain. \citet{2010ApJ...724..546S} studied direct impact accretion in eccentric orbits by treating transferred material as ballistic streams. They found that disk and direct impact accretion is not separated by a single $R_{\mathrm{min}}$ but a range, within which the accretion modes also depends on the orbital eccentricity and the initial donor rotation rate. Direct impact accretion has also been studied in the context of white dwarf accretion \citep{2014ApJ...785..157S,2015ApJ...806...76K} with the ballistic models of \citet{2010ApJ...724..546S}. \citet{2014ApJ...785..157S} found that angular momentum transported to the accretor depends on the binary properties because the angle for the particle to hit the star determines how much angular momentum is transported to the star and the orbit. 

In our model, we assume that a decretion disk forms around rapidly rotating stars for the direct impact accretion regime. We further assume that the transferred material carries a specific angular momentum $j_{\mathrm{tr}}=\sqrt{GM_{\mathrm{a}}1.7R_{\mathrm{min}}}$, corresponding to the specific angular momentum of the ballistic mass-transfer stream. A caveat of this treatment is that we neglect the detailed interaction between the stream and the disk, which could modify the effective angular momentum deposition. A more accurate description of this process would require dedicated multi-dimensional hydrodynamical simulations.

In the regime of direct impact accretion, it has been proposed that a hot spot that forms at the stellar surface where the accretion stream hits the accretor could result in non-conservative mass accretion if the luminosity of the hot spot exceeds the local Eddington luminosity \citep{2008A&A...487.1129V,2010A&A...510A..13V,2011A&A...528A..16V,2013A&A...557A..40D}. This mechanism could indicate an overestimate of mass accretion at low rotation rate in our model. Despite the reduced accretion, the star is still supposed to be spun up to critical rotation quickly \citep{2013A&A...557A..40D}. At the high rotation regime, \citet{2013A&A...557A..40D} assumed that a disk is formed whenever the star has reached $80\%$ of the critical rotation rate, which is necessary for removing angular momentum from the star to allow continued accretion. However, the existence of the disk could lead to high uncertainties in this model as the hot spot might form at the disk edge rather than the stellar surface.  

If the mass transfer rate is sufficiently high, the accretion flow may become geometrically thick near the outer edge of the disk, and the transferred material could acquire high enough energy to escape through $L_{2}$ point \citep{2023MNRAS.519.1409L,2025ApJ...990..172S,2025arXiv251024127S}. The threshold mass transfer rate at which mass loss through $L_{2}$ becomes significant is affected by factors such as the gas composition, the viscosity parameter, and mass ratio \citep{2023MNRAS.519.1409L}. In general, systems experiencing high mass transfer rates are expected to experience more non-conservative mass transfer. As a result, this effect should be more prominent in more massive binaries.                                                                                                                                                                                                                                                                                                                                                                                              


\subsection{Accretion efficiency inferred from other binaries}

Apart from Be star binaries, a variety of other binary systems and astrophysical processes have been used to interpret potential mass accretion efficiency between non-degenerate stars. For example, \citet{2007A&A...467.1181D} found that case~A detached and semi-detached systems in the Small Magellanic Cloud exhibit a wide range of accretion efficiencies. Similar conclusions have been drawn for massive Algol binaries in the Milky Way, Large and Small Magellanic Clouds, where accretion efficiencies span a broad range \citep{2022A&A...659A..98S, 2025arXiv251115347S}. Studies of Wolf-Rayet+O binaries imply that mass transfer is highly non-conservative \citep{2005A&A...435.1013P,2016ApJ...833..108S,2025A&A...695A.117N}. The companion star mass from type II supernova explosions put a constraint that does not support highly conservative mass transfer \citep{2025arXiv250812677Z}. These systems lie within diverse ranges of masses, mass ratios, and orbital configurations. The varied accretion efficiencies inferred both among different binary systems and within the same class demonstrates a complication in mass transfer that cannot be captured by a single efficiency value or by strictly rotationally or thermally limited prescriptions. Under the disk-star coupling model, the complexity may arise from disk formation conditions, variations in disk properties, the efficiency of mass loss through a hot spot \citep{2013A&A...557A..40D}, and mass loss from the outer disk boundary \citep{2023MNRAS.519.1409L}.

\section{Conclusion} \label{sec:conclusion}

Motivated by the discrepancy between the high binary accretion efficiencies inferred from Be+sdOB systems and the low efficiencies predicted by commonly adopted rotationally limited accretion model in binary simulations, we develop a mass-transfer prescription for rotating non-degenerate stars based on the idea that an accretion or decretion disk can regulate angular momentum of accreting stars, following the framework proposed by \citep{1991ApJ...370..604P}. We find that the accretor is spun up to near critical rotation rapidly due to mass accretion, and further accretion can still proceed despite the accretor's radial expansion further reducing the critical rotation rate, as long as the disk can extract the stellar angular momentum efficiently. The mass accretion efficiency is determined self-consistently from mass and angular momentum conservation in a quasi-steady disk.

We perform analytical calculations of accretion efficiencies from our model and show its relation to the stellar rotation rate, the ratio of the stellar radius to its Roche-lobe radius, and the binary mass ratio. The accretion efficiencies in our binary models are broadly consistent with the values expected near the adopted rotational threshold, where the specific angular momentum of the accreted material declines sharply. This is because the mass-weighted accretion is dominated by the phase of the most rapid mass transfer, during which efficient angular momentum exchange between the star and the disk is required to keep the star near the threshold rotation rate. The accretion efficiencies can be much higher than those from the commonly adopted rotationally limited accretion model in \texttt{MESA} simulations.

In this work, we focus on assessing whether our model can produce systems that occupy the region of parameter space where the observed Be+sdOB systems are found. We show that rotationally limited accretion model results in systematically low Be-star masses in our simulated Be and helium star binaries when compared with the observed Be+sdOB binaries, while the disk-star coupling accretion model can produce component masses in agreement with the observations for Be+sdOB binaries. Our discussion highlights that the complexity of accretion efficiency may stem from disk formation conditions, disk properties, tidal effects on the disk, and the extent of mass loss from the interaction between the accretion stream and the disk or the star. 

\begin{acknowledgments}
This work was supported by the Gordon and Betty Moore Foundation (PI Kalogera, grant award GBMF8477) and the Swiss National Science Foundation (PI Fragos, project number CRSII5\_213497). V.K. was partially supported through the D. I. Linzer Distinguished University Professorship fund. E.Z. acknowledges support from the Hellenic Foundation for Research and Innovation (H.F.R.I.) under the ``3rd Call for H.F.R.I. Research Projects to support Post-Doctoral Researchers" (Project No: 7933). This research was supported in part through the computational resources and staff contributions provided for the Quest high performance computing facility at Northwestern University which is jointly supported by the Office of the Provost, the Office for Research, and Northwestern University Information Technology.

\end{acknowledgments}

\appendix
\restartappendixnumbering
\section{Regime of disk accretion and direct impact accretion}

\begin{figure}
\includegraphics{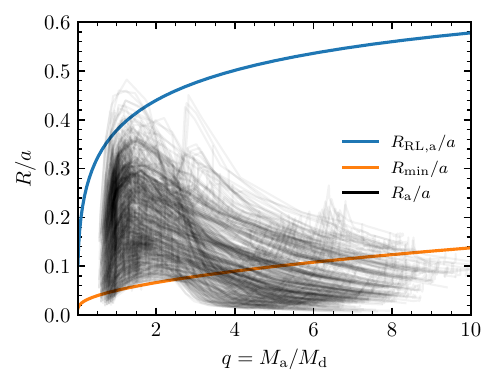}
\caption{$R_{\mathrm{min}}/a$ and $R_{\mathrm{RL,a}}/a$ as a function of mass ratio are shown in blue and orange line, respectively. The black lines show the evolution of $R_{\mathrm{a}}/a$ of the selected Be and helium star binaries in our grid under disk-star coupling model. 
\label{fig:r/a}}
\end{figure}

In this section, we present in Figure~\ref{fig:r/a} the ratios of $R_{\mathrm{min}}$ and $R_{\mathrm{RL,a}}$ over the orbital separation $a$ as a function of mass ratio $q=M_{\mathrm{a}}/M_{\mathrm{d}}$. We also show the evolutionary tracks of $R_{\mathrm{a}}/a$ of the selected Be and helium star binaries computed with the disk–star coupling model. For most of their evolution, these systems reside in the direct-impact accretion regime. Only after a substantial amount of mass has been transferred and the orbital separation has increased significantly do the majority of binaries transition into the disk-accretion regime.

\section{Model variations in accreted angular momentum} \label{appendixb}

\begin{figure}
\includegraphics{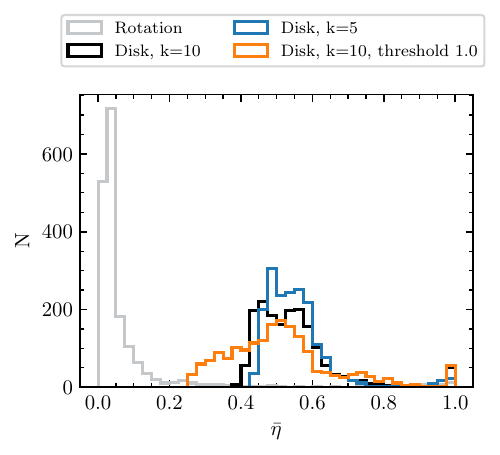}
\caption{Distribution of the mass-weighted accretion efficiency in our binary models for the different parameter variations in $j_{\mathrm{acc}}$.
\label{fig:a_eta_v}}
\end{figure}

In section~\ref{subsec:ama}, we discuss that the functional form adopted for $j_{\mathrm{acc}}$ has only a modest impact on the accretion efficiency. Here, we illustrate this by presenting the analytical accretion efficiency for direct impact accretion in Figure~\ref{fig:eta_v}, and the mass-weighted accretion efficiency for our binary grids in Figure~\ref{fig:a_eta_v}, for two variations of the $j_{\mathrm{acc}}$ prescription. 

The first variation adopts a lower value, $k=5$, instead of $k=10$. This cause $j_{\mathrm{acc}}$ to decline earlier and less steeply with increasing rotation rate. As a result, the analytical accretion efficiency is lower below the threshold rotation rate but higher above it. Since the accretion efficiency in the binary simulations is primarily determined by the narrow range of rotation rates just above the threshold, the mass-weighted accretion efficiencies obtained with $k = 5$ are very similar to those with $k = 10$, differing only by a slight increase. 

The second variation shifts the threshold rotation rate from $0.9$ to $1.0$. In this case, the mass-weighted accretion efficiency spans a broader range down to $\sim0.3$, because some binaries can reach rotation rates that overshoot to values exceeding unity, which enhances mass loss. Nevertheless, the mass-weighted accretion efficiency is peaking around $0.5$.

\textbf{\begin{figure*}[t]
\includegraphics[width=\textwidth]{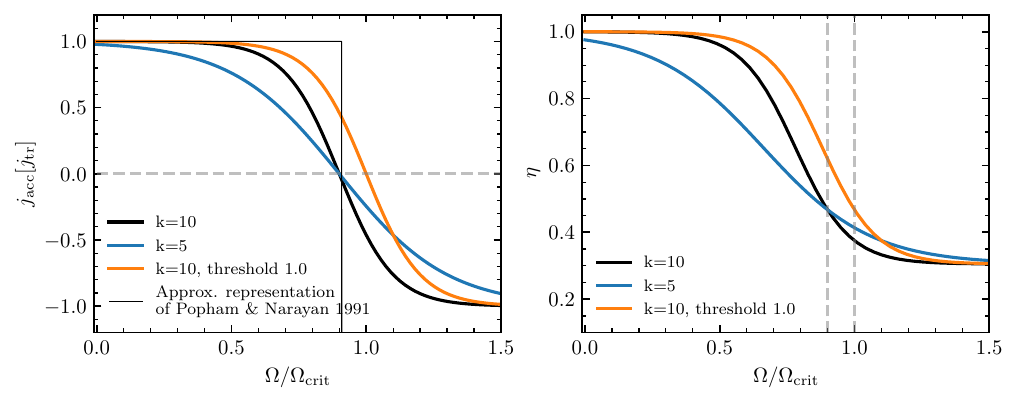}
\caption{Left panel: adopted dependence of $j_{\mathrm{acc}}$ on $\Omega/\Omega_{\mathrm{crit}}$ for $k=10$ with a threshold rotation rate of $0.9$, and two additional cases exploring parameter variations with $k=5$ and a threshold rotation rate of $1.0$, respectively. The dashed line denotes $j_{\mathrm{acc}}=0$. Right panel: corresponding analytical accretion efficiency as a function of $\Omega/\Omega_{\mathrm{crit}}$ for a mass ratio of $1.0$ in the direct-impact regime. The dashed lines mark the threshold values of $0.9$ and $1.0$.
\label{fig:eta_v}}
\end{figure*}}

\bibliography{main}{}
\bibliographystyle{aasjournalv7}



\end{document}